\documentclass[aps,prd,eqsecnum,preprint,tightenlines,nofootinbib,showpacs]{revtex4-1}
\usepackage{graphicx,amsmath,latexsym}

\newcommand{\bea}{\begin{eqnarray}}
\newcommand{\eea}{\end{eqnarray}}
\newcommand{\be}{\begin{equation}}
\newcommand{\ee}{\end{equation}}
\newcommand{\ub}[1]{\underline{#1}}
\newcommand{\ob}[1]{\overline{#1}}
\newcommand{\Pminus}{{\cal P}^-}
\newcommand{\veck}{\vec{k}_\perp}

\begin{document}

\title{A light-front coupled cluster method\footnote{Presented by J.R. Hiller
at LIGHTCONE 2011, 23-27 May 2011, Dallas.  To appear in the proceedings.}}

\author{John R. Hiller}
\author{Sophia S. Chabysheva}
\affiliation{Department of Physics \\
University of Minnesota-Duluth \\
Duluth, Minnesota 55812}

\date{\today}

\begin{abstract}
A new method for the nonperturbative solution
of quantum field theories is described.  
The method adapts the exponential-operator technique
of the standard many-body coupled-cluster method
to the Fock-space eigenvalue problem for light-front
Hamiltonians.  This leads to an effective eigenvalue problem
in the valence Fock sector and a set of nonlinear integral
equations for the functions that define the exponential 
operator.  The approach avoids at least some of the difficulties
associated with the Fock-space truncation usually used.
\end{abstract}

\maketitle

\section{Introduction}
\label{sec:intro}

\begin{figure}[b]
\centering
  \includegraphics[width=12cm]{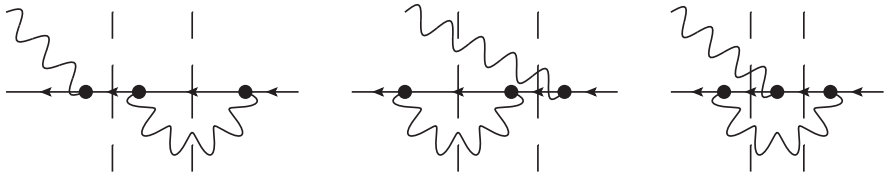}
\caption{Graphs that contribute to the Ward identity at lowest order.
The dashed vertical lines mark intermediate states.  Only the first
graph survives a truncation to Fock states with only one photon.}
\label{fig:fig1}
\end{figure}

Nonperturbative Hamiltonian methods that rely on Fock-space truncation
are plagued by uncanceled divergences.  The truncation removes contributions
from couplings to higher Fock sectors that would otherwise cancel against
contributions that are kept.  When these contributions separately diverge
with respect to an infinite-regulator-scale limit, an uncanceled divergence
is born.  For example, the Ward identity in QED requires contributions that
differ in the number of photons required, as illustrated in Fig.~\ref{fig:fig1};
a truncation in photon number will break the Ward identity~\cite{OSUQED} and
introduce an uncanceled divergence.

One approach to the handling of uncanceled divergences is to not take
the infinite-regulator limit and instead seek a range of regulator scales
over which physical quantities are slowly varying~\cite{OnePhotonQED}.
In this range, one has a compromise between errors due to truncation, which increase
as the regulator scale is increased, and errors due to the presence of the regulator,
which decrease as the scale is increased.  This approach has been used
in nonperturbative calculations of the anomalous magnetic moment of
the electron regulated by inclusion of Pauli--Villars (PV)
fields~\cite{OnePhotonQED,TwoPhotonQED}.

An alternative approach is sector-dependent renormalization~\cite{Wilson},
where the bare parameters are allowed to depend on the Fock sector.
This absorbs the uncanceled divergences into redefinitions of the
coupling constants.  Calculations of the anomalous moment have
been done this way~\cite{hb,Karmanov}.  However, the Fock-state wave
functions become ill-defined without proper normalization~\cite{SecDep},
and the bare coupling is driven to zero at finite regulator scales~\cite{hb}.
In fact, the sector-dependent charge renormalization is not ordinary
charge renormalization; it occurs in truncated QED without the presence of
fermion loops and is the result of wave function renormalization
in the broken Ward identity.

To avoid these difficulties we consider a new method, the light--front
coupled-cluster (LFCC) method~\cite{LFCClett}, that does
not involve Fock-space truncation.  It is instead based on the
exponential-operator technique of the standard many-body coupled-cluster (CC)
method~\cite{CCorigin} frequently employed in nuclear and chemical
physics~\cite{CCreviews}.  We construct an eigenstate of a
light-front Hamiltonian as $\sqrt{Z}e^T|\phi\rangle$, where
$|\phi\rangle$ is a valence state with a minimum number of
constituents, $T$ is an operator that increases particle number
without violating any conservation laws,
and $\sqrt{Z}$ is a normalization factor.  The truncation that
is made is in $T$ but not in the exponentiation, so that all
relevant Fock states are retained.  This infinite expansion
in Fock states necessitates some care in the computation 
of matrix elements, to avoid doing an infinite sum, but the
techniques of the standard CC method can be used, with some
extensions, to do the computations.

In the standard CC method $|\phi\rangle$ is a product of 
single-particle states and terms in $T$ annihilate states
in $|\phi\rangle$ and create excited states, to build in
correlations.  The operator $T$ is then truncated at 
some number of excitations, usually no more than four.
The focus is on finding the solution in the valence
sector with a large but fixed number of particles.
In contrast, in the LFCC method, the focus is on finding a
solution in the infinite Fock space that has the same quantum
numbers as the valence state $|\phi\rangle$.  The number
of particles in the valence state not large, and how
this state is to be determined is left unspecified.

Some applications to field theory of the CC method have been previously
considered~\cite{CC-QFT}, for Fock-state expansions in equal-time
quantization.  The focus is on the non-trivial vacuum structure,
which is avoided in light-front quantization, and
particle states are then built on the vacuum.  There was
some success in the analysis of $\phi^4_{1+1}$, particularly
of symmetry-breaking effects.

For the LFCC method, light-front quantization is crucial,
not only for the simple vacuum but also because the
internal motion of a coupled system can be factored 
from the external motion.  Of course, this separation
can be done in the nonrelativistic CC method but not
for a relativistic method in equal-time quantization.
Obviously, we then use light-cone coordinates~\cite{Dirac,DLCQreview},
which we define as $x^+=t+z$ for time, $\ub{x}=(x^-,\vec{x}_\perp)$ for
space, with $x^-\equiv t-z$ and $\vec{x}_\perp=(x,y)$.  The light-cone
energy is $p^-=E-p_z$, and the momentum is $\ub{p}=(p^+,\vec{p}_\perp)$,
with $p^+\equiv E+p_z$ and $\vec{p}_\perp=(p_x,p_y)$.  These bring
the mass-shell condition $p^2=m^2$ to the form $p^-=\frac{m^2+p_\perp^2}{p^+}$.
The Hamiltonian eigenvalue problem is
\be \label{eq:eigenvalueproblem}
\Pminus|\psi\rangle=\frac{M^2+P_\perp^2}{P^+}|\psi\rangle .
\ee

In the next Section, we give a more detailed 
description of the LFCC method.  This is followed
in Sec.~\ref{sec:QED} by a formulation of light-front
QED in an arbitrary gauge~\cite{ArbGauge}.  In Sec.~\ref{sec:application},
the two formulations are applied to a calculation
of the electron's anomalous moment.  A brief summary
is given in Sec.~\ref{sec:summary}.  Additional details
can be found elsewhere in these proceedings~\cite{ch}.

\section{LFCC method} \label{sec:LFCC}

We wish to solve the fundamental eigenvalue problem (\ref{eq:eigenvalueproblem})
and compute physical quantities from matrix elements between eigenstates.
The starting point is to write an eigenstate as $|\psi\rangle=\sqrt{Z}e^T|\phi\rangle$.
The factor $Z$ maintains the chosen normalization,
$\langle\psi'|\psi\rangle=\delta(\ub{P}'-\ub{P})$.  The valence state
$|\phi\rangle$ is also normalized: $\langle\phi'|\phi\rangle=\delta(\ub{P}'-\ub{P})$.
The operator $T$ contains terms that only increase particle number, such that it
conserves $J_z$, light-front momentum $\ub{P}$, charge, and any other conserved
quantum number.  Because $p^+$ is always positive, $T$ must include annihilation,
and, therefore, powers of $T$ include contractions.  These contractions will
introduce loop corrections.

To solve the eigenvalue problem, we construct an effective Hamiltonian
$\ob{{\cal P}-}\equiv e^{-T} \Pminus e^T$ and let $P_v$ project onto the
valence Fock sector.  With these definitions, we arrive at the
coupled system
\be
P_v\ob{\Pminus}|\phi\rangle=\frac{M^2+P_\perp^2}{P^+}|\phi\rangle,\;\; 
(1-P_v)\ob{\Pminus}|\phi\rangle=0 .
\ee
The second equation determines the operator $T$, which then fully
defines the effective Hamiltonian in the first equation.  The
two equations are essentially a coupled set of integral equations
for the valence state and the functions that define the terms in $T$.
In general, they must be solved by numerical methods.

The expectation value for an operator $\hat{O}$ is given by
\be
\langle\hat O\rangle=\frac{\langle\phi| e^{T^\dagger}\hat O e^T|\phi\rangle}
                      {\langle\phi| e^{T^\dagger} e^T|\phi\rangle} .
\ee
Direct computation requires infinite sums.  Following the techniques
of the CC method~\cite{CCreviews}, we instead define $\ob{O}=e^{-T}\hat O e^T$ and 
\be
\langle\tilde\psi|=\langle\phi|\frac{e^{T^\dagger}e^T}
      {\langle\phi|e^{T^\dagger} e^T|\phi\rangle} .
\ee
We then find that $\langle\hat O\rangle=\langle\tilde\psi|\ob{O}|\phi\rangle$  and
\be
\langle\tilde\psi'|\phi\rangle
=\langle\phi'|\frac{e^{T^\dagger}e^T}{\langle\phi| e^{T^\dagger} e^T|\phi\rangle}|\phi\rangle
=\delta(\ub{P}'-\ub{P}).
\ee
The effective operator $\ob{O}$ can be computed from its Baker--Hausdorff expansion
\be
\ob{O}=\hat O + [\hat O,T]+\frac12[[\hat O,T],T]+\cdots .
\ee
The bra $\langle\tilde\psi|$ is a left eigenvector of $\ob{\Pminus}$, as
can be seen from the following:
\be
\langle\tilde\psi|\ob{\Pminus}
=\langle\phi|\frac{e^{T^\dagger}\Pminus e^T}{\langle\phi| e^{T^\dagger} e^T|\phi\rangle}
=\langle\phi|\ob{\Pminus}^\dagger \frac{e^{T^\dagger}e^T}
                            {\langle\phi| e^{T^\dagger} e^T|\phi\rangle}
=\frac{M^2+P_\perp^2}{P^+}\langle\tilde\psi|.
\ee

We still have an infinite system of equations.  To arrive at a finite
set, we truncate $T$ at a fixed increase in particle count and
truncate the projection $(1-P_v)$ to a set of Fock sectors
above the valence sector that yield enough equations to solve for 
the functions in $T$.  This leaves a finite system of nonlinear
equations for the functions in $T$ and in the valence state.
The effective Hamiltonian $\ob{\Pminus}=e^{-T}\Pminus e^T$ has only
a finite number of terms in its Baker--Hausdorff expansion, 
$\ob{\Pminus} = \Pminus + [\Pminus,T] + \cdots $ and is computed
from these commutators of $T$ and $\Pminus$.  Similarly, the
expansion of an effective operator $\ob{O}$ has only a finite
number of terms.  The left-hand eigenstate
$|\tilde\psi\rangle$ is also truncated to a consistent
set of Fock sectors, no more than found in $T|\phi\rangle$.

\section{QED in an arbitrary covariant gauge} \label{sec:QED}

In order to facilitate a check of gauge invariance for physical
observables, we quantize light-front QED in an arbitrary covariant
gauge~\cite{ArbGauge}.  The PV-regulated Lagrangian is~\cite{TwoPhotonQED}
\be
{\cal L} =  \sum_{i=0}^2 (-1)^i \left[-\frac14 F_i^{\mu \nu} F_{i,\mu \nu} 
         +\frac12 \mu_i^2 A_i^\mu A_{i\mu} 
         -\frac12 \zeta \left(\partial^\mu A_{i\mu}\right)^2\right]
    + \sum_{i=0}^2 (-1)^i \bar{\psi_i} (i \gamma^\mu \partial_\mu - m_i) \psi_i 
  - e \bar{\psi}\gamma^\mu \psi A_\mu ,
\ee
where the fermion and photon fields $\psi_0$ and $A_{0\mu}$ are the 
physical fields and indices of 1 and 2 indicate PV fields.  The
interaction term is built from null combinations
\be \label{eq:NullFields}
  \psi =  \sum_{i=0}^2 \sqrt{\beta_i}\psi_i, \;\;
  A_\mu  = \sum_{i=0}^2 \sqrt{\xi_i}A_{i\mu} .
\ee
The coupling coefficients satisfy the constraints
\be
\xi_0=1, \;\;
\sum_{i=0}^2(-1)^i\xi_i=0, \;\;
\beta_0=1, \;\;
\sum_{i=0}^2(-1)^i\beta_i=0, \;\;
\ee
with $\xi_2$ and $\beta_2$ fixed by
chiral symmetry restoration~\cite{ChiralLimit,ArbGauge}
and zero photon mass~\cite{VacPol}.

To quantize, we apply a light-front analog of Stueckelberg
quantization~\cite{Stueckelberg}.  Consider the Lagrangian
of a free massive vector field:
${\cal L}=-\frac14 F^2+\frac12\mu A^2-\frac12\zeta(\partial\cdot A)^2$.
The field equation is $(\Box +\mu^2)A_\mu-(1-\zeta)\partial_\mu(\partial\cdot A)=0$,
and the light-front Hamiltonian density is
\be
{\cal H}={\cal H}|_{\zeta=1}
  +\frac12(1-\zeta)(\partial\cdot A)(\partial\cdot A
                                -2\partial_-A^+-2\partial_\perp\cdot\vec A_\perp),
\ee
with ${\cal H}|_{\zeta=1}=\frac12\sum_{\mu=0}^3 \epsilon^\mu
       \left[(\partial_\perp A^\mu)^2+\mu^2 (A^\mu)^2\right]$ 
and $\epsilon^\mu=(-1,1,1,1)$.  The field equation is satisfied
by
\be
A_\mu(x)=\int\frac{d\ub{k}}{\sqrt{16\pi^2 k^+}}\left\{\sum_{\lambda=1}^3
   e_\mu^{(\lambda)}(\ub{k})\left[ a_\lambda(\ub{k})e^{-ik\cdot x}
            + a_\lambda^\dagger(\ub{k})e^{ik\cdot x}\right]
+e_\mu^{(0)}(\ub{k})\left[ a_0(\ub{k})e^{-i\tilde k\cdot x}
            + a_0^\dagger(\ub{k})e^{i\tilde k\cdot x}\right]\right\}
\ee
with $\mu_\lambda=\mu$ for $\lambda=1,2,3$, 
but $\mu_0=\tilde\mu\equiv\mu/\sqrt{\zeta}$,
$\ub{\tilde k}=\ub{k}$, and $\tilde k^-=(k_\perp^2+\tilde\mu^2)/k^+$.
The polarization vectors
\be
e^{(1,2)}(\ub{k})=(0,2 \hat e_{1,2}\cdot \vec{k}_\perp/k^+,\hat e_{1,2}), \;\;
e^{(3)}(\ub{k})=\frac1\mu((k_\perp^2-\mu^2)/k^+,k^+,\vec k_\perp), \;\;
e^{(0)}(\ub{k})=\tilde k/\mu ,
\ee 
satisfy $k\cdot e^{(\lambda)}=0$ and 
$e^{(\lambda)}\cdot e^{(\lambda')}=-\delta_{\lambda\lambda'}$
for $\lambda,\lambda'=1,2,3$.
The first term in $A_\mu$ satisfies $(\Box +\mu^2)A_\mu=0$ and
$\partial\cdot A=0$.  The $\lambda=0$ term violates each, but 
the field equation is satisfied.
The nonzero commutators are
$[a_\lambda(\ub{k}),a_{\lambda'}^\dagger(\ub{k}')]
     =\epsilon^\lambda \delta_{\lambda\lambda'}\delta(\ub{k}-\ub{k}')$.
The resulting light-front Hamiltonian is
\be
\Pminus=\int d\ub{x}{\cal H}|_{x^+=0}
  =\int d\ub{k} \sum_\lambda \epsilon^\lambda \frac{k_\perp^2+\mu_\lambda^2}{k^+}
                a_\lambda^\dagger(\ub{k})a_\lambda(\ub{k}).
\ee

The nondynamical components of the fermion fields satisfy the
constraints ($i=0,1,2$)
\be \label{eq:FermionConstraint}
i(-1)^i\partial_-\psi_{i-}+e A_-\sqrt{\beta_i}\sum_j\psi_{j-}
  =(i\gamma^0\gamma^\perp)
     \left[(-1)^i\partial_\perp \psi_{i+}-ie A_\perp\sqrt{\beta_i}\sum_j\psi_{j+}\right] 
   -(-1)^i m_i \gamma^0\psi_{i+} .
\ee
When multiplied by $(-1)^i\sqrt{\beta_i}$ and summed over $i$, this becomes
\be
i\partial_-\psi_-
  =(i\gamma^0\gamma^\perp)
     \partial_\perp \psi_+
      - \gamma^0\sum_i\sqrt{\beta_i}m_i\psi_{i+},
\ee
which is the same constraint as for a free fermion field, in any gauge.

Thus, the light-front QED Hamiltonian can be constructed in any covariant
gauge, to obtain
\bea
\Pminus &=&   \sum_{i,s}\int d\ub{p}
      \frac{m_i^2+p_\perp^2}{p^+}(-1)^i
          b_{i,s}^\dagger(\ub{p}) b_{i,s}(\ub{p})
+\sum_{l,\lambda}\int d\ub{k}
          \frac{\mu_{l\lambda}^2+k_\perp^2}{k^+}(-1)^l\epsilon^\lambda
             a_{l\lambda}^\dagger(\ub{k}) a_{l\lambda}(\ub{k})  \\
&& +\sum_{ijls\sigma\lambda}\int dy d\vec{k}_\perp 
   \int\frac{d\ub{p}}{\sqrt{16\pi^3p^+}}
             \left\{h_{ijl}^{\sigma s\lambda}(y,\vec{k}_\perp)\right.  \nonumber \\
&& \left. \times  a_{l\lambda}^\dagger(yp^+,y\vec{p}_\perp+\vec{k}_\perp)
   b_{js}^\dagger((1-y)p^+,(1-y)\vec{p}_\perp-\vec{k}_\perp)b_{i\sigma}(\ub{p})  + \mbox{H.c.}\right\}.  \nonumber
\eea
The functions $h_{ijl}^{\sigma s\lambda}$ are readily computed from spinor matrix
elements.  Terms with positron contributions have been neglected, for simplicity.

\section{Application of the LFCC method} \label{sec:application}

We briefly discuss an application to the calculation of the
anomalous magnetic moment of the electron.  For simplicity,
Fock space is truncated to exclude positrons.  The valence state
is $|\phi^\pm(\ub{P})\rangle=\sum_i z_i b_{i\pm}^\dagger(\ub{P})|0\rangle$.
The $T$ operator is truncated to be
\be
T=\sum_{ijls\sigma\lambda}\int dy d\vec{k}_\perp 
   \int\frac{d\ub{p}}{\sqrt{16\pi^3}}\sqrt{p^+} t_{ijl}^{\sigma s\lambda}(y,\vec{k}_\perp)
 a_{l\lambda}^\dagger(yp^+,y\vec{p}_\perp+\vec{k}_\perp)
   b_{js}^\dagger((1-y)p^+,(1-y)\vec{p}_\perp-\vec{k}_\perp)b_{i\sigma}(\ub{p}) .
\ee
After exclusion of terms that annihilate the valence state,
the effective Hamiltonian is
\bea
\ob{\Pminus} &=& \sum_{ijs}\int d\ub{p}(-1)^i
      \left[\delta_{ij}\frac{m_i^2+p_\perp^2}{p^+}+\frac{I_{ji}}{p^+}\right]
          b_{j,s}^\dagger(\ub{p}) b_{i,s}(\ub{p})   \\
&& +\sum_{ijls\sigma\lambda}\int dy d\vec{k}_\perp 
   \int\frac{d\ub{p}}{\sqrt{16\pi^3p^+}}
          \left\{h_{ijl}^{\sigma s\lambda}(y,\vec{k}_\perp)
            +\frac12 V_{ijl}^{\sigma s\lambda}(y,\vec k_\perp)\right.  \nonumber \\
&& +\left[\frac{m_j^2+k_\perp^2}{1-y}
                                +\frac{\mu_{l\lambda}^2+k_\perp^2}{y}-m_i^2\right]
                                t_{ijl}^{\sigma s\lambda}(y,\vec k_\perp) \nonumber \\
&& \left.
    +\frac12\sum_{i'}\frac{I_{ji'}}{1-y}t_{ii'l}^{\sigma s\lambda}(y,\vec k_\perp)
              -\sum_{j'}(-1)^{i+j'}t_{j'jl}^{\sigma s\lambda}(y,\vec k_\perp)I_{j'i}\right\}
                             \nonumber \\
&& \times  a_{l\lambda}^\dagger(yp^+,y\vec{p}_\perp+\vec{k}_\perp)
   b_{js}^\dagger((1-y)p^+,(1-y)\vec{p}_\perp-\vec{k}_\perp)b_{i\sigma}(\ub{p}), \nonumber
\eea
with the self-energy $I_{ji}$ and the vertex loop correction $V_{ijl}^{\sigma s\lambda}$
defined in \cite{ch}.

The projection 
$P_v \ob{\Pminus}|\phi^\pm(\ub{P})\rangle=\frac{M^2+P_\perp^2}{P^+}|\phi^\pm(\ub{P})\rangle$ 
of the eigenvalue problem yields
\be
m_i^2 z_{ai}^\pm +\sum_j I_{ij} z_{aj}^\pm = M_a^2 z_{ai}^\pm.
\ee
Projection onto $|e\gamma\rangle$, orthogonal to $|\phi\rangle$, gives
\be
\left[M_a^2-\frac{M_b^2+k_\perp^2}{1-y}-\frac{\mu_{l\lambda}^2+k_\perp^2}{y}\right]
   C_{abl}^{\pm s\lambda}(y,\veck)
  =H_{abl}^{\pm s\lambda}(y,\veck)
   +\frac12\left[V_{abl}^{\pm s\lambda}(y,\veck)
      -\sum_{b'}\frac{I_{bb'}}{1-y}C_{ab'l}^{\pm s\lambda}(y,\veck)\right],
\ee
with
\be
C_{abl}^{\pm s\lambda}(y,\veck)
  \equiv\sum_{ij}(-1)^{i+j}z_{ai}^\pm \tilde{z}_{bj}^\pm t_{ijl}^{\pm s\lambda}(y,\veck).
\ee
Notice that $C_{abl}^{\pm s\lambda}$, the analog of the two-body wave function,
satisfies an equation where the bare masses have been replaced by the physical
masses $M_a$.  Thus, the LFCC method provides a natural way for physical masses
to enter the calculation instead of having them imposed by a sector-dependent
renormalization.  Also, although the $T$ operator has been truncated to a single
photon emission, the projected eigenvalue problem retains the self-energy
corrections and vertex corrections necessary for the Ward identity to be
satisfied.

\section{Summary} \label{sec:summary}

Details of the application to QED can be found in \cite{ch}.  The
result for the anomalous moment agrees with the Schwinger
correction at first order and will sum to all orders in $\alpha$
many of the higher-order corrections.

The key advantage of the LFCC method is that it avoids Fock-space
truncations that can induce uncanceled divergences and other
inconsistencies, such as Fock-sector dependence and spectator
dependence.  The approach is systematically improvable, by
the addition of terms to the truncated $T$ operator.

The complete calculation of the dressed-electron state
and its anomalous moment, within the given truncation of $T$,
is in progress.  It requires a numerical solution of the
right-hand and left-hand eigenvalue problems, as well as
numerical quadrature for the matrix element that yields
the anomalous moment.  Work beyond this can include
investigation of the dressed-photon state, extension
of the dressed-electron state to include $e^+e^-$ pairs, and
study of muonium, positronium, and symmetry breaking in
scalar theories.  Of particular interest, of course,
is the adaptation of these methods to QCD; the LFCC method
is general enough that it
can be applied once a suitably regulated light-front
Hamiltonian is constructed.

\acknowledgments
This work was supported in part by the US Department of Energy
through Contract No.\ DE-FG02-98ER41087
and by the Minnesota Supercomputing Institute through
grants of computing time.

\end{document}